\documentclass[twocolumn,showpacs,superscriptaddress,preprintnumbers,amsmath,amssymb]{revtex4}

\usepackage{graphicx}
\usepackage{dcolumn}
\usepackage{bm}

\newcommand{\bq}{\begin{equation}}
\newcommand{\eq}{\end{equation}}
\newcommand{\bqa}{\begin{eqnarray}}
\newcommand{\eqa}{\end{eqnarray}}
\newcommand{\nn}{\nonumber \\}

\def\be     {\begin{equation}}
\def\ee     {\end{equation}}
\def\bea        {\begin{eqnarray}}
\def\eea        {\end{eqnarray}}
\def\bnn    {\begin{eqnarray*}}
\def\enn    {\end{eqnarray*}}

\begin{document}

\title{Superconductivity from purely repulsive interactions
in the strong coupling approach : Application of the SU(2)
slave-rotor theory to the Hubbard model}
\author{Ki-Seok Kim}

\affiliation{Asia Pacific Center for Theoretical Physics, Hogil
Kim Memorial building 5th floor, POSTECH, Hyoja-dong, Namgu,
Pohang 790-784, Korea}

\author{Mun Dae Kim}

\affiliation{Institute of Physics and Applied Physics, Yonsei
University, Seoul 120-749, Korea}

\date{\today}

\begin{abstract}
We propose a mechanism of superconductivity from purely repulsive
interactions in the strong coupling regime, where the BCS
(Bardeen-Cooper-Schrieffer) mechanism such as the spin-fluctuation
approach is difficult to apply. Based on the SU(2) slave-rotor
representation of the Hubbard model, we find that the single
energy scale for the amplitude formation of Cooper pairs and their
phase coherence is separated into two energy scales, allowing the
so called pseudogap state where such Cooper pairs are coherent
locally but not globally, interpreted as realization of the
density-phase uncertainty principle. This superconducting state
shows the temperature-linear decreasing ratio of superfluid
weight, resulting from strong phase fluctuations.
\end{abstract}


\maketitle

\section{Introduction}

To find the mechanism of superconductivity from purely repulsive
interactions has been one of the central interests during the last
two decades associated with high $T_{c}$ cuprates
\cite{Review_SC}. It was shown that purely repulsive interactions
can turn into attractive ones via some renormalization processes
associated with spin-density-wave fluctuations, if the
Fermi-surface topology has a special nesting structure as  the
case of high T$_{c}$ cuprates \cite{HighTc} or Fe-based
superconductors \cite{FeSC}, for example, described by the so
called spin-fluctuation approach. In this approach phonons are
replaced with antiferromagnetic fluctuations taking the role of
pairing glue \cite{Spin_Fermion}, thus basically the same as the
strong coupling BCS (Bardeen-Cooper-Schrieffer) theory in the
Eliashberg approximation \cite{Eliashberg}. Unfortunately, such an
approach loses its theoretical validity in the strong coupling
regime because the spin-fermion model itself and its evaluation
way are justified only in the weak coupling limit \cite{FMQCP},
i.e., $U/D \ll 1$ with the interaction strength $U$ and half
bandwidth $D$.

This reminds us of two kinds of theories for magnetism; the
Hertz-Moriya-Millis (HMM) theory is the standard framework for
itinerant electrons \cite{HMM} while the magnetism from localized
spins is successfully described by the Schwinger-boson gauge
theory \cite{Schwinger_Boson}. Importantly, the strong coupling
approach of the Schwinger-boson theory gives rise to two kinds of
energy scales, associated with formation of short range
antiferromagnetic correlations and long range ordering for
antiferromagnetism. Emergence of the spin-gapped state above an
antiferromagnetic order reflects strong quantum fluctuations of
spin dynamics, guaranteed by the uncertainty relation of spins.

In this paper we propose a mechanism of superconductivity in the
strong coupling regime. Recently, one of us has suggested an SU(2)
slave-rotor representation of the Hubbard model, where not only
local density fluctuations but also on-site pairing excitations
are taken into account on equal footing, giving rise to
superconducting fluctuations naturally \cite{Kim_Rotor}. We here
obtain an effective U(1) gauge theory called the pair-rotor theory
of the Hubbard model, where phase fluctuations of Cooper pairs are
extracted from the SU(2) slave-rotor description. We show that the
single energy scale for the amplitude formation of Cooper pairs
and their phase coherence is separated into two energy scales,
allowing the so called pseudogap state where the Cooper pairs are
coherent locally but not globally. One can say that this
superconducting state resembles the RVB (resonating-valance-bond)
superconductivity \cite{RVB}. However the present mechanism is of
charge-fluctuation-induced, while for the RVB the slave-boson
study of the t-J model \cite{SU2SB} is of
spin-fluctuation-induced, where charge fluctuations are completely
frozen out.

Superconductivity of the U(1) pair-rotor theory is analogous with
antiferromagnetism of the Schwinger-boson theory, where the
amplitude formation of Cooper pairs and their global phase
coherence correspond to short range antiferromagnetic correlations
and condensation of Schwinger bosons, respectively, and the
density-phase uncertainty matches with the uncertainty relation
between spins. In this respect the pseudogap state of the
pair-rotor theory agrees with the spin-gapped phase of the
Schwinger-boson theory, identified with the hallmark of the strong
coupling approach. On the other hand, this description should be
differentiated from the U(1) slave-rotor theory of the t-J-U model
where d-wave singlet pairs originate from the spin-exchange term,
but the U(1) slave-rotor field has nothing to do with pairing
fluctuations \cite{Kim_Rotor}.

We would like to point out an interesting reformulation of the
Hubbard model, where the role of on-site pairing fluctuations is
emphasized \cite{Phillips}. Although this formulation differs from
the present SU(2) slave-rotor theory which is a gauge theory, it
also suggests that such quantum fluctuations give rise to
superconducting correlations in doped Mott insulators.

\section{SU(2) slave-rotor representation of the Hubbard model}

We rewrite the Hubbard model \bqa && H = - t\sum_{\langle ij
\rangle }(c_{i\sigma}^{\dagger}e^{iA_{ij}}c_{j\sigma} + H.c.) +
\frac{3u}{2}\sum_{i}c_{i\uparrow}^{\dagger}c_{i\uparrow}
c_{i\downarrow}^{\dagger}c_{i\downarrow}
\nn \eqa as follows \cite{Kim_Rotor}, \bqa && Z = \int{D[\eta_{i},
U_{i}, \vec{\Omega}_{i}, E_{ij},
F_{ij}]}e^{-\int_{0}^{\beta}{d\tau} L} , \nn && L = L_{0} +
L_{\eta} + L_{U} , ~~~ L_{0} = t\sum_{\langle ij
\rangle}\mathbf{tr}(F_{ij}E_{ij}^{\dagger} + H.c.) , \nn &&
L_{\eta} = \sum_{i}\eta_{i}^{\dagger}(\partial_{\tau}\mathbf{I} -
i\vec{\Omega}_{i}\cdot\vec{\tau})\eta_{i} - t\sum_{\langle ij
\rangle}(\eta_{i}^{\dagger}F_{ij}\eta_{j} + H.c.) , \nn && L_{U} =
\frac{1}{4u}\sum_{i}\mathbf{tr}(-iU_{i}\partial_{\tau}U_{i}^{\dagger}
+ \vec{\Omega}_{i}\cdot\vec{\tau} +
i\mu{U}_{i}\tau_{3}U_{i}^{\dagger})^{2} \nn && - t\sum_{\langle ij
\rangle}\mathbf{tr}(U_{j}^{\dagger}E_{ij}^{\dagger}U_{i}e^{iA_{ij}\tau_{3}}\tau_{3}
+ H.c.) ,  \eqa where an electron $\Phi_{i} = \left(\begin{array}{c} c_{i\uparrow} \\
c_{i\downarrow}^{\dagger} \end{array} \right)$ is assumed to be a
composite of a holon $U_{i} = \left( \begin{array}{cc} z_{i\uparrow} & - z_{i\downarrow}^{\dagger} \\
z_{i\downarrow} & z_{i\uparrow}^{\dagger} \end{array} \right)$ and
a spinon $\eta_{i}= \left(\begin{array}{c} \eta_{i+} \\
\eta_{i-}^{\dagger} \end{array} \right)$ carrying charge and spin
quantum numbers, respectively, given by \bqa && \Phi_{i} =
U_{i}^{\dagger} \eta_{i} \eqa with the constraint
$|z_{i\uparrow}|^{2} + |z_{i\downarrow}|^{2} = 1$. $E_{ij}$ and
$F_{ij}$ are 2$\times$2 matrix fields associated with hopping of
holons and spinons, respectively, and $\vec{\Omega}_{i}$ is an
isospin field related with on-site density and pairing potentials.
$\mu$ is an electron chemical potential, and $A_{ij}$ is an
external electromagnetic field.

It is not difficult to see equivalence between the SU(2)
slave-rotor effective Lagrangian [Eq. (2)] and Hubbard model [Eq.
(1)], integrating over field variables of $E_{ij}$, $F_{ij}$ and
$\vec{\Omega}_{i}$ and replacing the composite field
$U_{i}^{\dagger} \eta_{i}$ with an electron field $\Phi_{i}$. The
procedure is well described in the previous study
\cite{Kim_Rotor}. An important feature in the SU(2) slave-rotor
description is the emergence of pairing correlations between
nearest neighbor electrons, given by off diagonal hopping in
$F_{ij}$ which results from on-site pairing fluctuations, captured
by the off diagonal variable $z_{i\downarrow}$ of the SU(2) matrix
field $U_{i}$.
However, the appearance of pairing correlations does not
necessarily lead to superconductivity because their global
coherence, described by condensation of SU(2) matrix holons, is
not guaranteed. The similar situation happens in the SU(2)
slave-boson theory of the t-J model \cite{SU2SB}.

We write SU(2) hopping matrices as $E_{ij} \approx E W_{ij}
\tau_{3}$ and $F_{ij} \approx F W_{ij} \tau_{3}$, where their
amplitudes are assumed to be homogeneous and the SU(2) phase
factor can be represented as \bqa && W_{ij} \equiv \left(
\begin{array}{cc} X_{ij} & - Y_{ij}^{\dagger} \\
Y_{ij} & X_{ij}^{\dagger} \end{array} \right) \nonumber \eqa
without losing generality, satisfying the unitary constraint
$|X_{ij}|^{2} + |Y_{ij}|^{2} = 1$. Then, we find an effective
SU(2) slave-rotor action \bqa && Z = \int{D[\eta_{is},
z_{i\sigma}, \vec{\Omega}_{i}, X_{ij}, Y_{ij}]}
\delta(|X_{ij}|^{2} + |Y_{ij}|^{2} - 1) \nn &&
\delta(|z_{i\uparrow}|^{2} + |z_{i\downarrow}|^{2} - 1) \exp\Bigl(
-\int_{0}^{\beta}{d\tau} L \Bigr) , \nn && L = L_{\eta} + L_{U} +
4t\sum_{\langle ij \rangle} EF , \nn && L_{\eta} = \sum_{i} \left(
\begin{array}{cc} \eta_{i+}^{\dagger} & \eta_{i-}
\end{array} \right) \left(
\begin{array}{cc} \partial_{\tau} - i \Omega_{i}^{z} & - i[\Omega_{i}^{x} - i \Omega_{i}^{y}] \\
- i[\Omega_{i}^{x} + i \Omega_{i}^{y}] & \partial_{\tau} + i
\Omega_{i}^{z} \end{array} \right) \nn && \left( \begin{array}{c} \eta_{i+} \\
\eta_{i-}^{\dagger} \end{array} \right) - tF\sum_{\langle ij
\rangle}\Bigl\{\left( \begin{array}{cc} \eta_{i+}^{\dagger} &
\eta_{i-} \end{array} \right) \left(
\begin{array}{cc} X_{ij} & Y_{ij}^{\dagger} \\
Y_{ij} & - X_{ij}^{\dagger} \end{array} \right) \left( \begin{array}{c} \eta_{j+} \\
\eta_{j-}^{\dagger} \end{array} \right) \nn && + H.c. \Bigr\} ,
\nn && L_{U} = \frac{1}{4u}\sum_{i} \Bigl\{ \Bigl( -
i[z_{i\uparrow}\partial_{\tau} z_{i\uparrow}^{\dagger} +
z_{i\downarrow}^{\dagger} \partial_{\tau} z_{i\downarrow}] +
\Omega_{i}^{z} \nn && + i \mu (|z_{i\uparrow}|^{2} -
|z_{i\downarrow}|^{2} + 1) \Bigr)^{2} + \Bigl( -
i[z_{i\uparrow}\partial_{\tau} z_{i\downarrow}^{\dagger} -
z_{i\downarrow}^{\dagger}\partial_{\tau} z_{i\uparrow}] \nn && +
\Omega_{i}^{x} - i \Omega_{i}^{y} + 2 i\mu
z_{i\uparrow}z_{i\downarrow}^{\dagger} \Bigr) \Bigl( - i[
z_{i\downarrow}\partial_{\tau} z_{i\uparrow}^{\dagger}  -
z_{i\uparrow}^{\dagger} \partial_{\tau} z_{i\downarrow}] \nn && +
\Omega_{i}^{x} + i \Omega_{i}^{y} + 2 i\mu
z_{i\uparrow}^{\dagger}z_{i\downarrow}\Bigr) + \Bigl( - i[
z_{i\downarrow}\partial_{\tau} z_{i\uparrow}^{\dagger}  -
z_{i\uparrow}^{\dagger} \partial_{\tau} z_{i\downarrow}] \nn &&  +
\Omega_{i}^{x} + i \Omega_{i}^{y} + 2 i\mu
z_{i\uparrow}^{\dagger}z_{i\downarrow}\Bigr) \Bigl( -
i[z_{i\uparrow}\partial_{\tau} z_{i\downarrow}^{\dagger} -
z_{i\downarrow}^{\dagger}\partial_{\tau} z_{i\uparrow}] \nn && +
\Omega_{i}^{x} - i \Omega_{i}^{y} + 2 i\mu
z_{i\uparrow}z_{i\downarrow}^{\dagger} \Bigr) + \Bigl(- i[
z_{i\downarrow}\partial_{\tau} z_{i\downarrow}^{\dagger} +
z_{i\uparrow}^{\dagger}\partial_{\tau} z_{i\uparrow} ] \nn && -
\Omega_{i}^{z} - i \mu (|z_{i\uparrow}|^{2} -
|z_{i\downarrow}|^{2} + 1) \Bigr)^{2} \Bigr\} \nn && -
2tE\sum_{\langle ij \rangle}\Bigl\{\left( \begin{array}{cc}
z_{i\uparrow}^{\dagger} & z_{i\downarrow}^{\dagger} \end{array}
\right) \left(
\begin{array}{cc} X_{ij}e^{-iA_{ij}} & Y_{ij}^{\dagger}e^{-iA_{ij}} \\
Y_{ij}e^{-iA_{ij}} & - X_{ij}^{\dagger}e^{-iA_{ij}} \end{array}
\right) \nn && \left( \begin{array}{c} z_{j\uparrow} \\
z_{j\downarrow} \end{array} \right) + H.c. \Bigr\} .  \eqa

Our main problem is how to extract dynamics for phase fluctuations
of pairing order parameters from the boson sector of the SU(2)
slave-rotor theory. The easy axis approximation of $U_{i} =
e^{i\phi_{i}\tau_{3}}$ implying $z_{i\downarrow} = 0$ does not
allow pairing correlations, identified with on-site density
fluctuations and giving rise to the Mott transition from a
paramagnetic Mott insulator to a Fermi liquid metal via their
condensation \cite{U1SR}. In this study we take an easy plane
limit, introducing pairing correlations. Justification of this
approximation can be given in the similar way as the SU(2)
slave-boson theory \cite{SU2SB}.

\section{U(1)-pair slave-rotor theory}

\subsection{Easy plane approximation}

We introduce an isospin field \bqa && \vec{I}_{i} \equiv
\frac{1}{2} z_{i\sigma}^{\dagger} \vec{\tau}_{\sigma\sigma'}
z_{i\sigma'} , \nonumber \eqa and consider an easy plane limit
\bqa && \vec{I}_{i} = I_{i}^{x} \hat{x} + I_{i}^{y} \hat{y}
\nonumber \eqa with $I_{i}^{x 2} + I_{i}^{y 2} = 1/2$, described
by \bqa && z_{i\uparrow} = \frac{1}{\sqrt{2}}
e^{i\phi_{i\uparrow}} , ~~~~~ z_{i\downarrow} = \frac{1}{\sqrt{2}}
e^{i\phi_{i\downarrow}} . \eqa Inserting Eq. (5) into Eq. (4), we
find \bqa && L_{\phi} = \frac{1}{2u}\sum_{i} \Bigl\{
\Bigl(\frac{1}{2}[
\partial_{\tau} \phi_{i\uparrow} -
\partial_{\tau}\phi_{i\downarrow}] - \Omega_{i}^{z} - i \mu \Bigr)^{2}
\nn && + \Bigl( \frac{1}{2}[\partial_{\tau} \phi_{i\uparrow} +
\partial_{\tau} \phi_{i\downarrow}] - [\Omega_{i}^{x} + i
\Omega_{i}^{y}] e^{i(\phi_{i\uparrow} - \phi_{i\downarrow})} -
i\mu  \Bigr) \nn && \Bigl(\frac{1}{2}[
\partial_{\tau} \phi_{i\uparrow} + \partial_{\tau}
\phi_{i\downarrow}] - [\Omega_{i}^{x} - i \Omega_{i}^{y}] e^{-
i(\phi_{i\uparrow} - \phi_{i\downarrow})} - i\mu \Bigr) \Bigr\}
\nn && - t E\sum_{\langle ij \rangle}\Bigl\{\left(
\begin{array}{cc} e^{-i\phi_{i\uparrow}} &
e^{-i\phi_{i\downarrow}} \end{array} \right) \left(
\begin{array}{cc} X_{ij}e^{-iA_{ij}} & Y_{ij}^{\dagger}e^{-iA_{ij}} \\
Y_{ij}e^{-iA_{ij}} & - X_{ij}^{\dagger}e^{-iA_{ij}} \end{array}
\right) \nn &&
\left( \begin{array}{c} e^{i\phi_{j\uparrow}} \\
e^{i\phi_{j\downarrow}} \end{array} \right) + H.c. \Bigr\} ,  \eqa
where the spinon part is the same as that of Eq. (4). As shown in
this effective theory, the presence of the off diagonal term in
the kinetic energy of rotorons allows us to control global phase
coherence of spinon-pairing excitations. We note that this
effective Lagrangian is analogous with that of the $d-wave$
pairing state in the SU(2) slave-boson theory, where gauge
fluctuations may give rise to composite pairing fluctuations
between different boson species \cite{SU2SB}, corresponding to
$(\phi_{i\uparrow} + \phi_{i\downarrow})/2$ in the U(1) pair-rotor
theory.

The isospin field was argued to prefer an easy plane in the
non-linear $\sigma$ model description of the SU(2) slave-boson
theory when holes are doped, resulting from an effective potential
for the easy plane anisotropy \cite{SU2SB}. Even if the easy plane
approximation is difficult to justify self-consistently, the
present formulation gives us a chance to investigate the role of
pairing fluctuations beyond the conventional description.

\subsection{Gauge transformation}

\subsubsection{Spinon sector}

One can make the phase factor gauged away in the phase-gauge
coupling term of Eq. (6), performing the gauge transformation \bqa
&& \Omega_{i}^{x} - i \Omega_{i}^{y} \rightarrow (\Omega_{i}^{x} -
i \Omega_{i}^{y}) e^{ i (\phi_{i\uparrow} - \phi_{i\downarrow})} ,
\nn && \Omega_{i}^{x} + i \Omega_{i}^{y} \rightarrow
(\Omega_{i}^{x} + i \Omega_{i}^{y}) e^{- i (\phi_{i\uparrow} -
\phi_{i\downarrow})} . \eqa Then, the spinon Lagrangian in Eq. (4)
is given by \bqa && L_{\eta} = \sum_{i} \left(
\begin{array}{cc} \eta_{i+}^{\dagger} & \eta_{i-}
\end{array} \right) \nn && \left(
\begin{array}{cc} \partial_{\tau} - i \Omega_{i}^{z} &
- i(\Omega_{i}^{x} - i \Omega_{i}^{y}) e^{ i (\phi_{i\uparrow} -
\phi_{i\downarrow})} \\
- i(\Omega_{i}^{x} + i \Omega_{i}^{y}) e^{- i (\phi_{i\uparrow} -
\phi_{i\downarrow})} & \partial_{\tau} + i
\Omega_{i}^{z} \end{array} \right) \nn && \left( \begin{array}{c} \eta_{i+} \\
\eta_{i-}^{\dagger} \end{array} \right) - tF\sum_{\langle ij
\rangle}\Bigl\{\left( \begin{array}{cc} \eta_{i+}^{\dagger} &
\eta_{i-} \end{array} \right) \left(
\begin{array}{cc} X_{ij} & Y_{ij}^{\dagger} \\
Y_{ij} & - X_{ij}^{\dagger} \end{array} \right) \left( \begin{array}{c} \eta_{j+} \\
\eta_{j-}^{\dagger} \end{array} \right) \nn && + H.c. \Bigr\} .
\nonumber \eqa

To make the phase factor gauged away in the off diagonal part of
the spinon Lagrangian, we introduce the gauge transformation of
\bqa && \psi_{i\uparrow} = e^{- i (\phi_{i\uparrow} -
\phi_{i\downarrow})/2} \eta_{i+} , \nn &&
\psi_{i\downarrow}^{\dagger} = e^{ i (\phi_{i\uparrow} -
\phi_{i\downarrow})/2} \eta_{i-}^{\dagger} , \eqa where
$\psi_{i\sigma}$ is a renormalized spinon via virtual pairing
fluctuations. Then, the phase field appears in the time derivative
and SU(2) gauge matrix $W_{ij}$. Considering the gauge
transformation \bqa && \Omega_{i}^{z} \rightarrow \Omega_{i}^{z} +
\frac{1}{2}(
\partial_{\tau}\phi_{i\uparrow} -
\partial_{\tau} \phi_{i\downarrow}) , \nn && X_{ij} \rightarrow e^{i (\phi_{i\uparrow} -
\phi_{i\downarrow})/2} X_{ij} e^{- i (\phi_{j\uparrow} -
\phi_{j\downarrow})/2} , \nn && Y_{ij} \rightarrow e^{- i
(\phi_{i\uparrow} - \phi_{i\downarrow})/2} Y_{ij} e^{- i
(\phi_{j\uparrow} - \phi_{j\downarrow})/2} ,  \eqa we find an
effective Lagrangian for renormalized spinons \bqa && L_{\psi} =
\sum_{i} \left( \begin{array}{cc} \psi_{i\uparrow}^{\dagger} &
\psi_{i\downarrow} \end{array} \right) \left(
\begin{array}{cc} \partial_{\tau} - i \Omega_{i}^{z} & - (\Omega_{i}^{x} - i \Omega_{i}^{y}) \\
- (\Omega_{i}^{x} + i \Omega_{i}^{y}) & \partial_{\tau} + i
\Omega_{i}^{z} \end{array} \right) \nn && \left( \begin{array}{c} \psi_{i\uparrow}  \\
\psi_{i\downarrow}^{\dagger}
\end{array} \right) - tF\sum_{\langle ij
\rangle}\Bigl\{ \left( \begin{array}{cc}
\psi_{i\uparrow}^{\dagger} & \psi_{i\downarrow} \end{array}
\right) \left(
\begin{array}{cc} X_{ij} & Y_{ij}^{\dagger} \\ Y_{ij} & - X_{ij}^{\dagger} \end{array} \right) \left(
\begin{array}{c} \psi_{j\uparrow}  \\ \psi_{j\downarrow}^{\dagger}
\end{array} \right) \nn && + H.c. \Bigr\} , \nonumber \eqa where the phase
field is removed completely.

\subsubsection{Pairon sector}

Based on Eqs. (7) and (9), we obtain \bqa && L_{\phi} =
\frac{1}{2u}\sum_{i} \Bigl\{ \Bigl( \Omega_{i}^{z} + i \mu
\Bigr)^{2} + \Bigl( \frac{1}{2}[\partial_{\tau} \phi_{i\uparrow} +
\partial_{\tau} \phi_{i\downarrow}] \nn && - [\Omega_{i}^{x} + i
\Omega_{i}^{y}] - i\mu \Bigr) \Bigl(\frac{1}{2}[
\partial_{\tau} \phi_{i\uparrow} + \partial_{\tau}
\phi_{i\downarrow}] - [\Omega_{i}^{x} - i \Omega_{i}^{y}] \nn && -
i\mu \Bigr) \Bigr\} - 2 tE\sum_{\langle ij \rangle} \Bigl\{ \Bigl(
Y_{ij}^{\dagger} + Y_{ij} \Bigr) \cos \Bigl(
\frac{\phi_{i\uparrow}+\phi_{i\downarrow}}{2} \nn && -
\frac{\phi_{j\uparrow}+\phi_{j\downarrow}}{2} + A_{ij} \Bigr) + i
\Bigl( X_{ij}^{\dagger} - X_{ij} \Bigr) \sin \Bigl(
\frac{\phi_{i\uparrow}+\phi_{i\downarrow}}{2} \nn && -
\frac{\phi_{j\uparrow}+\phi_{j\downarrow}}{2} + A_{ij} \Bigr)
\Bigr\} . \nonumber \eqa Introducing new phase variables \bqa &&
\frac{1}{2}[\phi_{i\uparrow}+\phi_{i\downarrow}] \equiv \phi_{ic}
, ~~~~~ \frac{1}{2}[\phi_{i\uparrow}-\phi_{i\downarrow}] \equiv
\phi_{is} , \eqa the rotor Lagrangian becomes \bqa && L_{\phi} =
\frac{1}{2u}\sum_{i} \Bigl\{ \Bigl( \Omega_{i}^{z} + i \mu
\Bigr)^{2} \nn && + \Bigl( \partial_{\tau} \phi_{ic} -
[\Omega_{i}^{x} + i \Omega_{i}^{y}] - i\mu \Bigr) \Bigl(
\partial_{\tau} \phi_{ic} - [\Omega_{i}^{x} - i \Omega_{i}^{y}] - i\mu
\Bigr) \Bigr\} \nn && - 2 tE\sum_{\langle ij \rangle} \Bigl\{
\Bigl( Y_{ij}^{\dagger} + Y_{ij} \Bigr) \cos \Bigl( \phi_{ic} -
\phi_{jc} + A_{ij}\Bigr) \nn && + i \Bigl( X_{ij}^{\dagger} -
X_{ij} \Bigr) \sin \Bigl( \phi_{ic} - \phi_{jc} + A_{ij} \Bigr)
\Bigr\} , \nonumber \eqa where anomalous phase-gauge couplings are
gauged away. Comparing this pairon Lagrangian with Eq. (6), we see
that the phase field of $\phi_{is}$ disappears via gauge
transformation. We call $\phi_{ic}$ pairon because it controls
coherence of local singlet pairs, basically the same role as
Schwinger-bosons in the Schwinger-boson theory
\cite{Schwinger_Boson}.

\subsection{U(1) pair-rotor effective Lagrangian}

We write down an effective U(1) pair-rotor theory of the Hubbard
model for phase-fluctuating superconductivity \bqa && Z =
\int{D[\psi_{i\sigma}, \phi_{ic}, \vec{\Omega}_{i}, X_{ij},
Y_{ij}]} \delta(|X_{ij}|^{2} + |Y_{ij}|^{2} - 1) \nn && \exp
\Bigl( -\int_{0}^{\beta}{d\tau} L \Bigr) , ~~~ L = L_{\psi} +
L_{\phi} + 4t\sum_{\langle ij \rangle} EF , \nn && L_{\psi} =
\sum_{i}\psi_{i\alpha}^{\dagger}(\partial_{\tau}
\delta_{\alpha\beta} -
i\vec{\Omega}_{i}\cdot\vec{\tau}_{\alpha\beta})\psi_{i\beta} \nn
&& - tF\sum_{\langle ij \rangle}
(\psi_{i\alpha}^{\dagger}W_{ij}^{\alpha\beta}\tau_{3}\psi_{j\beta}
+ H.c.) , \nn && L_{\phi} = \frac{1}{2u}\sum_{i} \Bigl(
\partial_{\tau} \phi_{ic} - [\Omega_{i}^{x} + i \Omega_{i}^{y}] -
i\mu \Bigr) \nn && \Bigl(
\partial_{\tau} \phi_{ic} - [\Omega_{i}^{x} - i \Omega_{i}^{y}] - i\mu
\Bigr) \nn && - 2 tE\sum_{\langle ij \rangle} \Bigl\{ \Bigl(
Y_{ij}^{\dagger} + Y_{ij} \Bigr) \cos \Bigl( \phi_{ic} - \phi_{jc}
+ A_{ij}\Bigr) \nn && + i \Bigl( X_{ij}^{\dagger} - X_{ij} \Bigr)
\sin \Bigl( \phi_{ic} - \phi_{jc} + A_{ij} \Bigr) \Bigr\} \nn && +
\frac{1}{2u}\sum_{i} ( \Omega_{i}^{z} + i \mu )^{2} ,  \eqa where
superconductivity is characterized by condensation of pairons
$\langle e^{i\phi_{ic}} \rangle \not= 0$ in the presence of
pairing correlations, $Y_{ij}$, basically the same as the fact
that antiferromagnetism of localized spins is described by
condensation of Schwinger-bosons in the presence of local
antiferromagnetic correlations. Actually, the pairon field is
identified with the phase field of the pairing order parameter,
since $Y_{ij}$ plays the role of phase stiffness for $\phi_{ic}$.

An interesting feature of the U(1) pair-rotor theory is emergence
of two energy scales from the single energy scale $u/t$ in the
Hubbard model, corresponding to appearance of incoherent singlet
correlations and global coherence of such preformed pairs. The
former energy scale may be identified with the pseudogap
temperature $T^{*}$, and the latter will be the superconducting
transition temperature $T_{c}$. Considering that the spinon sector
is nothing but the BCS theory in the mean-field approximation,
$T^{*}$ is expected to coincide with the mean-field transition
temperature of the BCS theory. On the other hand, the pairon
Lagrangian corresponds to the XY model in the mean-field
approximation, thus $T_{c}$ will be the coherence temperature of
the XY model.

\subsection{Electron as a composite of spinon and pairon}

An electron field can be represented as a composite object of a
spinon and a pairon, \bqa && \left( \begin{array}{c} c_{i\uparrow} \\
c_{i\downarrow}^{\dagger} \end{array} \right)
= \frac{1}{\sqrt{2}} \left( \begin{array}{cc} e^{- i \phi_{ic}} & e^{- i \phi_{ic}} \\
- e^{ i \phi_{ic}} & e^{ i \phi_{ic}} \end{array} \right)
\left( \begin{array}{c} \psi_{i\uparrow} \\
\psi_{i\downarrow}^{\dagger} \end{array} \right) , \eqa where
condensation of pairons $\langle e^{-i\phi_{ic}} \rangle \not= 0$
recovers the BCS quasiparticle relation, allowing
superconductivity. Inserting the U(1) pair-rotor representation
[Eq. (12)] into the Hubbard model [Eq. (1)], one can obtain the
U(1) pair-rotor theory [Eq. (11)] from the Hubbard model directly.
The inverse transformation expresses the spinon field in terms of
a pairon field and an electron field, \bqa && \psi_{i\uparrow} =
\frac{1}{\sqrt{2}}e^{i \phi_{ic}}c_{i\uparrow} -
\frac{1}{\sqrt{2}}e^{- i \phi_{ic}}c_{i\downarrow}^{\dagger} , \nn
&& \psi_{i\downarrow}^{\dagger} = \frac{1}{\sqrt{2}}e^{ i
\phi_{ic}} c_{i\uparrow} + \frac{1}{\sqrt{2}}e^{ - i \phi_{ic}}
c_{i\downarrow}^{\dagger} . \nonumber \eqa

Using Eq. (12), one can write down the Cooper pair field as \bqa
&& \Delta_{ij}^{cp} \equiv \langle c_{i\uparrow} c_{j\downarrow} -
c_{i\downarrow} c_{j\uparrow} \rangle \approx \frac{1}{2} \langle
e^{- i (\phi_{ic}+\phi_{jc}) } \rangle \nn && \Bigl( \langle
\psi_{i\uparrow}^{\dagger}\psi_{j\uparrow} +
\psi_{j\uparrow}^{\dagger} \psi_{i\uparrow} \rangle + \langle
\psi_{i\downarrow}^{\dagger}\psi_{j\downarrow} +
\psi_{j\downarrow}^{\dagger}\psi_{i\downarrow} \rangle \nn && +
\langle \psi_{i\uparrow}^{\dagger}\psi_{j\downarrow}^{\dagger} -
\psi_{i\downarrow}^{\dagger}\psi_{j\uparrow}^{\dagger} \rangle +
\langle \psi_{i\uparrow}\psi_{j\downarrow} -
\psi_{i\downarrow}\psi_{j\uparrow} \rangle \Bigr) , \nonumber \eqa
where not only particle-particle pairing of spinons but also their
particle-hole pairing is included. In this respect the pairing
symmetry of Cooper pairs has always an $s$-component although the
particle-particle channel is $d-wave$. However, this quantity
should not be considered to represent the true pairing symmetry of
the superconducting pair. Actually, it is measured from the
electron spectral function as an excitation gap, given by \bqa &&
G_{ij,\uparrow\uparrow}^{el} \equiv - \langle c_{i\uparrow}
c_{j\uparrow}^{\dagger} \rangle \nn && \approx - \frac{1}{2}
\langle e^{- i (\phi_{ic} - \phi_{jc}) } \rangle \langle
\psi_{i\uparrow}\psi_{j\uparrow}^{\dagger} +
\psi_{i\downarrow}^{\dagger}\psi_{j\uparrow}^{\dagger} +
\psi_{i\uparrow}\psi_{j\downarrow} +
\psi_{i\downarrow}^{\dagger}\psi_{j\downarrow} \rangle . \nn
\nonumber \eqa In this respect the pairing symmetry of the
superconducting order parameter will be $d-wave$ as far as the
spinon pairing order parameter is $d-wave$.

\subsection{$d-wave$ mean-field ansatz}

We take the $d-wave$ ansatz for the pairing field
$(Y_{ii+\hat{x}}, Y_{ii+\hat{y}}) = (Y, - Y)$ and uniform
approximation for the hopping parameter $(X_{ii+\hat{x}},
X_{ii+\hat{y}}) = (X, X)$. The pairing potential is set
$\Omega_{i}^{x,y} = 0$ in the mean-field approximation because
only virtual fluctuations ($z_{i\sigma}$) are allowed due to high
energy cost, while the density potential is replaced with
$\Omega_{i}^{z} = - i \varphi$ for notational convenience.
Introducing $b_{ic} = e^{i\phi_{ic}}$ with the rotor constraint
$|b_{ic}|^{2} = 1$, we write down the mean-field Lagrangian of the
U(1) pair-rotor theory in the momentum space, \bqa && Z =
\int{D[\psi_{k\sigma}, b_{kc} ]} e^{-\int_{0}^{\beta}{d\tau} L} ,
\nn && L = L_{\psi} + L_{\phi} + 8 N t E F + N \lambda_{g} (X^{2}
+ Y^{2} - 1)  , \nn && L_{\psi} = \sum_{k} \left(
\begin{array}{cc} \psi_{k\uparrow}^{\dagger} & \psi_{-k\downarrow}
\end{array} \right) \left(
\begin{array}{cc} \partial_{\tau} - \varphi & 0 \\
0 & \partial_{\tau} + \varphi  \end{array} \right) \left( \begin{array}{c} \psi_{k\uparrow}  \\
\psi_{-k\downarrow}^{\dagger}
\end{array} \right) \nn && - 2 tF\sum_{k} \left( \begin{array}{cc}
\psi_{k\uparrow}^{\dagger} & \psi_{-k\downarrow} \end{array}
\right) \left(
\begin{array}{cc} X \gamma_{k} & Y \varphi_{k} \\ Y \varphi_{k} & - X \gamma_{k} \end{array} \right) \left(
\begin{array}{c} \psi_{k\uparrow}  \\ \psi_{-k\downarrow}^{\dagger}
\end{array} \right)   ,
\nn && L_{U} = \frac{1}{2u} \sum_{k} [(i\partial_{\tau} + i \mu)
b_{kc}]^{2} - 4 t E Y\sum_{k} \varphi_{k} b_{kc}^{\dagger} b_{kc}
\nn && + \lambda_{c} \sum_{k} (|b_{kc}|^{2} - 1) + \frac{1}{2u}
\sum_{i} ( i \varphi - i \mu )^{2} , \eqa where $\gamma_{k} = \cos
k_{x} + \cos k_{y}$ and $\varphi_{k} = \cos k_{x} - \cos k_{y}$.
$N$ is number of lattice sites. $\lambda_{g}$ and $\lambda_{c}$
are Lagrange multiplier fields to impose the constraints for SU(2)
gauge-matrix fields and pair-rotor fields, respectively.

Performing integration of spinon and pairon fields, we find the
U(1) pair-rotor mean-field free energy \bqa && F[b, Y, E, F,
\lambda_{c}, \varphi, \mu ; \delta, T] = - \frac{2}{\beta}
\sum_{k} \ln \Bigl\{ 2 \cosh \Bigl( \frac{\beta E_{k}^{f}}{2}
\Bigr) \Bigr\} \nn && + \frac{1}{\beta} \sum_{q} \Bigl[ \ln
\Bigl\{ 2 \sinh \Bigl(\frac{\beta}{2} [\mathcal{E}_{q}^{b} - \mu]
\Bigr) \Bigr\} \nn && + \ln \Bigl\{ 2 \sinh \Bigl(\frac{\beta}{2}
[\mathcal{E}_{q}^{b} + \mu] \Bigr) \Bigr\} \Bigr] + N \Bigl( 8 t E
F - \frac{1}{2u} [\varphi - \mu]^{2} \nn && + \lambda_{c} [b^{2} -
1] - \frac{\mu^{2} }{2u} b^{2} - 8 t E Y b^{2} - \mu \delta \Bigr)
, \eqa where $b$, $\delta$, and $\beta$ are condensation
amplitude, hole concentration, and inverse temperature $1/T$,
respectively. The fermion spectrum \bqa && E_{k}^{f} = \sqrt{[2 t
F \sqrt{1-Y^{2}} \gamma_{k} + \varphi]^{2} + [2 t F Y
\varphi_{k}]^{2}} \nonumber \eqa coincides with the $d-wave$ BCS
theory \cite{dBCS}, and the boson spectrum is also relativistic,
\bqa && \mathcal{E}_{q}^{b} = \sqrt{- 8 u t E Y \varphi_{q} + 2 u
\lambda_{c}} , \nonumber \eqa basically the same as the
Schwinger-boson theory \cite{Schwinger_Boson}.

\subsection{Phase diagram}

It is interesting to observe that the pairon sector of the U(1)
pair-rotor theory is almost the same as the Schwinger-boson part
of the U(1) slave-fermion theory \cite{Jia_Kim}, where pairing
correlations or antiferromagnetic fluctuations give rise to
dynamics of pairons or Schwinger bosons, respectively. Actually,
we find that the pairing order parameter $Y$ decreases
monotonically as hole concentration increases. Since $Y$ acts as
the stiffness parameter for $b$, the condensation probability
$b^{2}$ is reduced (inset of Fig. 1). This is basically the same
as the slave-fermion theory \cite{Jia_Kim} where weakening of
antiferromagnetic correlations results in reduction of boson
condensation. On the other hand, the superconducting transition
temperature $T_{c}$, determined by vanishment of superfluid
density, is shown to increase as hole concentration increases in
small doping (Fig. 1).

\begin{figure}[t]
\vspace{5cm} \includegraphics{TcTsXB_MD.eps} \caption{ As hole concentration
$\delta$ increases, the superconducting transition temperature
$T_{c}(\delta)$ also increases in underdoped region ($u/t = 0.3$)
although the condensation amplitude $b^{2}(\delta)$ (inset)
decreases. } \label{fig1}
\end{figure}

\section{Superfluid density}

\subsection{Ioffe-Larkin composition rule}

We start from the U(1) pair-rotor theory \bqa && L_{f} = \sum_{i}
\left( \begin{array}{cc} \psi_{i\uparrow}^{\dagger} &
\psi_{i\downarrow} \end{array} \right) \left(
\begin{array}{cc} \partial_{\tau} - i a_{i\tau} & - i c_{i\tau}^{+} \\
- i c_{i\tau}^{-} & \partial_{\tau} + i a_{i\tau} \end{array}
\right) \left( \begin{array}{c} \psi_{i\uparrow}  \\
\psi_{i\downarrow}^{\dagger} \end{array} \right) \nn && -
tF\sum_{\langle ij \rangle}\Bigl\{ \left( \begin{array}{cc}
\psi_{i\uparrow}^{\dagger} & \psi_{i\downarrow} \end{array}
\right) \left( \begin{array}{cc} X e^{ia_{ij}} & Y e^{-ic_{ij}} \\
Y e^{ic_{ij}} & - X e^{-ia_{ij}} \end{array} \right) \left(
\begin{array}{c} \psi_{j\uparrow}  \\ \psi_{j\downarrow}^{\dagger}
\end{array} \right) \nn && + H.c. \Bigr\} ,
\nn && L_{b} = \frac{1}{2u}\sum_{i} ( \partial_{\tau} \phi_{ic} -
c_{i\tau}^{-} - i\mu ) ( \partial_{\tau} \phi_{ic} - c_{i\tau}^{+}
- i\mu ) \nn && - 2 tE\sum_{\langle ij \rangle} \Bigl\{ 2Y
\cos(c_{ij}) \cos \Bigl( \phi_{ic} - \phi_{jc} + A_{ij} \Bigr) \nn
&& - 2X \sin(a_{ij} ) \sin \Bigl( \phi_{ic} - \phi_{jc} + A_{ij}
\Bigr) \Bigr\} \nn && + \frac{1}{2u} \sum_{i} ( a_{i\tau} + i \mu
)^{2} , \eqa where low energy fluctuations of mean-field order
parameters are allowed, given by two kinds of gauge fields \bqa &&
X_{ij} = X e^{ia_{ij}} , ~~~ Y_{ij} = Y e^{ic_{ij}} \nonumber \eqa
for their spatial components and \bqa && a_{i\tau} =
\Omega_{i}^{z} , ~~~ c_{i\tau}^{\pm} = \Omega_{i}^{x} \mp i
\Omega_{i}^{y} \nonumber \eqa for their time components.

The partition function can be evaluated as a function of an
electromagnetic field, expanding the effective action up to the
second order for two kinds of gauge fluctuations, \bqa && Z_{A} =
\int D a_{ij} D c_{ij} D \psi_{i\sigma} D \phi_{ic} e^{-
\int_{0}^{\beta} d \tau (L_{f} + L_{b})} \nn && \approx \int D
a_{ij} D c_{ij} \exp\Bigl[ - F_{MF}^{f} - F_{MF}^{b} -
\frac{1}{2}\Bigl( \frac{\partial^{2} F_{f}}{\partial a_{ij}^{2}}
a_{ij}^{2} \nn && + 2 \frac{\partial^{2} F_{f}}{\partial a_{ij}
c_{ij}} a_{ij}c_{ij} + \frac{\partial^{2} F_{f}}{\partial
c_{ij}^{2}} c_{ij}^{2} \Bigr) - \frac{1}{2}\Bigl(
\frac{\partial^{2} F_{b}}{\partial a_{ij}^{2}} a_{ij}^{2} +
\frac{\partial^{2} F_{b}}{\partial c_{ij}^{2}} c_{ij}^{2} \nn && +
\frac{\partial^{2} F_{b}}{\partial A_{ij}^{2}} A_{ij}^{2} + 2
\frac{\partial^{2} F_{b}}{\partial a_{ij} c_{ij}} a_{ij}c_{ij} + 2
\frac{\partial^{2} F_{b}}{\partial a_{ij} A_{ij}} a_{ij}A_{ij} \nn
&& + 2 \frac{\partial^{2} F_{b}}{\partial c_{ij} A_{ij}}
c_{ij}A_{ij} \Bigr) \Bigr] , \eqa where \bqa &&
F_{f}[a_{ij},c_{ij}] = - \frac{1}{\beta} \ln \int D\psi_{i\sigma}
e^{-\int_{0}^{\beta} d \tau L_{f}[\psi_{i\sigma},a_{ij},c_{ij}]} ,
\nn && F_{b}[a_{ij},c_{ij},A_{ij}] = - \frac{1}{\beta} \ln \int
D\phi_{ic} e^{-\int_{0}^{\beta} d \tau
L_{b}[\phi_{ic},a_{ij},c_{ij},A_{ij}]} \nn \nonumber \eqa and \bqa
&& F_{MF}^{f} = F_{f}[0,0] , ~~~~~ F_{MF}^{b} = F_{b}[0,0,0] .
\nonumber \eqa

Performing the Gaussian integration for the two gauge fields, we
find the partition function with an electromagnetic field \bqa &&
Z_{A} \propto e^{ - F_{MF}^{f} - F_{MF}^{b}} \exp\Bigl[ -
\frac{1}{2} \Bigl\{ - \pi^{b}_{AA} + \frac{\pi^{b
2}_{aA}}{\pi^{f}_{aa} + \pi^{b}_{aa}} \nn && +
\frac{\Bigl(\pi^{b}_{cA} + \frac{\pi^{f}_{ac} +
\pi^{b}_{ac}}{\pi^{f}_{aa} + \pi^{b}_{aa}}
\Bigr)^{2}}{\pi^{f}_{cc} + \pi^{b}_{cc} - \frac{(\pi^{f}_{ac} +
\pi^{b}_{ac})^{2}}{\pi^{f}_{aa} + \pi^{b}_{aa}}} \Bigr\}
A_{ij}^{2} \Bigr] , \nonumber \eqa where $\pi_{\alpha\beta}^{f,b}
\equiv - (\partial^{2} F_{f,b})/(\partial \alpha \partial \beta)$
are current-current correlation functions with $\alpha, \beta = a,
c, A$. As a result, the superfluid density is given by \bqa &&
\rho_{s} = - \pi^{b}_{AA} + \frac{\pi^{b 2}_{aA}}{\pi^{f}_{aa} +
\pi^{b}_{aa}} + \frac{\Bigl(\pi^{b}_{cA} + \frac{\pi^{f}_{ac} +
\pi^{b}_{ac}}{\pi^{f}_{aa} + \pi^{b}_{aa}}
\Bigr)^{2}}{\pi^{f}_{cc} + \pi^{b}_{cc} - \frac{(\pi^{f}_{ac} +
\pi^{b}_{ac})^{2}}{\pi^{f}_{aa} + \pi^{b}_{aa}}} . \nn \eqa

\subsection{Each current-current correlation function}

The current-current correlation function can be derived as follows
\bqa && \pi^{b}_{AA} \equiv - \frac{\partial^{2}
F_{b}[A_{ij}]}{\partial A_{ij}^{2}} \nn && = \frac{1}{Z_{b} } \int
D\phi_{ic} \Bigl( - \frac{\partial S_{b}}{\partial A_{ij}} \Bigr)
\Bigl( - \frac{\partial S_{b}}{\partial A_{ij}} \Bigr) e^{-S_{b}}
\nn && - \Bigl\{ \frac{1}{Z_{b}} \int D\phi_{ic} \Bigl( -
\frac{\partial S_{b}}{\partial A_{ij} } \Bigr) e^{-S_{b}}
\Bigr\}^{2} \nn && + \frac{1}{Z_{b} } \int D\phi_{ic} \Bigl( -
\frac{\partial^{2} S_{b}}{\partial A_{ij}^{2} } \Bigr) e^{-S_{b}}
\nn && \equiv \langle j_{ij}^{bA} j_{ij}^{bA} \rangle - \langle
j_{ij}^{bA} \rangle^{2} + \langle K_{ij}^{bAA} \rangle . \eqa The
first and second terms show the paramagnetic response, given by
the current-current correlation function, while the last term
displays the diamagnetic response, expressed by the kinetic-energy
term. The "off diagonal" current-response function is given by
\bqa && \pi^{b}_{aA} \equiv - \frac{\partial^{2}
F_{b}[a_{ij},A_{ij}]}{\partial A_{ij}
\partial a_{ij} } \nn && = \frac{1}{Z_{b} } \int D\phi_{ic} \Bigl( - \frac{\partial
S_{b}}{\partial A_{ij}} \Bigr) \Bigl( - \frac{\partial
S_{b}}{\partial a_{ij}} \Bigr) e^{-S_{b}} \nn && - \Bigl\{
\frac{1}{Z_{b} } \int  D\phi_{ic} \Bigl( - \frac{\partial
S_{b}}{\partial A_{ij} } \Bigr) e^{-S_{b}} \Bigr\} \Bigl\{
\frac{1}{Z_{b} } \int D\phi_{ic} \Bigl( - \frac{\partial
S_{b}}{\partial a_{ij} } \Bigr) e^{-S_{b}} \Bigr\} \nn && +
\frac{1}{Z_{b} } \int D\phi_{ic} \Bigl( - \frac{\partial^{2}
S_{b}}{\partial A_{ij} \partial a_{ij} } \Bigr) e^{-S_{b}}  \nn &&
\equiv \langle j_{ij}^{bA} j_{ij}^{ba} \rangle - \langle
j_{ij}^{bA} \rangle \langle j_{ij}^{ba} \rangle + \langle
K_{ij}^{b aA} \rangle , \eqa basically the same as the above. All
other current-response functions are obtained in the same way as
this.

Currents and kinetic terms are \bqa && j_{ij}^{bA} \equiv -
\frac{\partial S_{b}}{\partial A_{ij}} = - 4 t E Y \sin \Bigl(
\phi_{ic} - \phi_{jc} \Bigr) , \nn && K_{ij}^{bAA} \equiv -
\frac{\partial^{2} S_{b}}{\partial A_{ij}^{2} } = - 4 t E Y \cos
\Bigl( \phi_{ic} - \phi_{jc} \Bigr)  , \nn && j_{ij}^{ba} \equiv -
\frac{\partial S_{b}}{\partial a_{ij}} = - 4 t E X \sin \Bigl(
\phi_{ic} - \phi_{jc} \Bigr) , \nn && K_{ij}^{baa} \equiv -
\frac{\partial^{2} S_{b}}{\partial a_{ij}^{2} } = 0 , ~~~
j_{ij}^{bc} \equiv - \frac{\partial S_{b}}{\partial c_{ij}} = 0 ,
\nn && K_{ij}^{bcc} \equiv - \frac{\partial^{2} S_{b}}{\partial
c_{ij}^{2} } = - 4 t E Y \cos \Bigl( \phi_{ic} - \phi_{jc} \Bigr)
, \nn && K_{ij}^{b aA} \equiv - \frac{\partial^{2} S_{b}}{\partial
A_{ij} \partial a_{ij} } = - 4 t E X \cos \Bigl( \phi_{ic} -
\phi_{jc} \Bigr)   , \nn && K_{ij}^{b cA} \equiv -
\frac{\partial^{2} S_{b}}{\partial A_{ij}
\partial c_{ij} } = 0 , \nn && K_{ij}^{b ac} \equiv -
\frac{\partial^{2} S_{b}}{\partial a_{ij}
\partial c_{ij} } = 0   \eqa for pairons and
\bqa &&  j_{ij}^{fa} \equiv - \frac{\partial S_{f}}{\partial
a_{ij}} = - i t F X (\psi_{i\sigma}^{\dagger}\psi_{j\sigma} -
\psi_{j\sigma}^{\dagger}\psi_{i\sigma}) , \nn && K_{ij}^{faa}
\equiv - \frac{\partial^{2} S_{f}}{\partial a_{ij}^{2}} = - t F X
(\psi_{i\sigma}^{\dagger}\psi_{j\sigma} +
\psi_{j\sigma}^{\dagger}\psi_{i\sigma}) , \nn && j_{ij}^{fc}
\equiv - \frac{\partial S_{f}}{\partial c_{ij}} = - i t F Y
(\psi_{i\uparrow}^{\dagger}\psi_{j\downarrow}^{\dagger} +
\psi_{i\downarrow}^{\dagger} \psi_{j\uparrow}^{\dagger}) \nn && -
i t F Y ( \psi_{i\downarrow}\psi_{j\uparrow} + \psi_{i\uparrow}
\psi_{j\downarrow} )   , \nn && K_{ij}^{fcc} \equiv -
\frac{\partial^{2} S_{f}}{\partial c_{ij}^{2}} = - t F Y
(\psi_{i\uparrow}^{\dagger}\psi_{j\downarrow}^{\dagger} -
\psi_{i\downarrow}^{\dagger} \psi_{j\uparrow}^{\dagger}) \nn && -
t F Y ( \psi_{i\downarrow}\psi_{j\uparrow} - \psi_{i\uparrow}
\psi_{j\downarrow} )  , \nn && K_{ij}^{f ac} \equiv -
\frac{\partial^{2} S_{f}}{\partial a_{ij} \partial c_{ij} } = 0
\eqa for spinons in equilibrium of $a_{ij}, c_{ij}, A_{ij}
\rightarrow 0$.

\subsection{Simplification in the expression of superfluid density}

Evaluating each correlation function, we find \bqa && \pi_{ac}^{f}
= 0 , ~~~~~ \pi_{ac}^{b} = 0 , ~~~~~ \pi_{cA}^{b} = 0 . \eqa It is
clear both mathematically and physically that these contributions
should vanish. Correlations between normal and pairing currents do
not exist in the spinon dynamics. Pairing-type currents do not
appear in the pairon sector, causing the second and third
equalities. As a result, the expression for the superfluid density
is simplified as follows \bqa && \rho_{s} = - \pi^{b}_{AA} +
\frac{\pi^{b 2}_{aA}}{\pi^{f}_{aa} + \pi^{b}_{aa}} , \eqa similar
to the conventional Ioffe-Larkin-type composition \cite{IL}.

\subsection{Evaluation of superfluid density}

Correlation functions for superfluid density are \bqa
\pi^{b}_{AA}(q,i\Omega) && = \langle j_{bA}(q,i\Omega)
j_{bA}(-q,-i\Omega) \rangle  \nn && - \langle j_{bA}(q,i\Omega)
\rangle \langle j_{bA}(-q,-i\Omega) \rangle + \langle
K_{AA}^{b}(q,i\Omega) \rangle , \nn \pi^{b}_{aA}(q,i\Omega) && =
\langle j_{ba}(q,i\Omega) j_{bA}(-q,-i\Omega) \rangle \nn && -
\langle j_{ba}(q,i\Omega) \rangle \langle j_{bA}(-q,-i\Omega)
\rangle + \langle K_{aA}^{b}(q,i\Omega) \rangle , \nn
\pi^{b}_{aa}(q,i\Omega) && = \langle j_{ba}(q,i\Omega)
j_{ba}(-q,-i\Omega) \rangle \nn && - \langle j_{ba}(q,i\Omega)
\rangle \langle j_{ba}(-q,-i\Omega) \rangle , \nn
\pi^{f}_{aa}(q,i\Omega) && = \langle j_{fa}(q,i\Omega)
j_{fa}(-q,-i\Omega) \rangle \nn && - \langle j_{fa}(q,i\Omega)
\rangle \langle j_{fa}(-q,-i\Omega) \rangle + \langle
K_{aa}^{f}(q,i\Omega) \rangle \nn  \eqa in the energy-momentum
space, where corresponding currents and kinetic energies are given
by \bqa && j_{bA}^{x}(q,i\Omega) = - 4 t E Y \sum_{k}
\sin\Bigl(k_{x} + \frac{q_{x}}{2}\Bigr) b_{kc}^{\dagger}b_{k+qc} ,
\nn && K_{AA}^{bx}(q,i\Omega) = - 4 t E Y \sum_{k} \cos\Bigl(k_{x}
+ \frac{q_{x}}{2}\Bigr) b_{kc}^{\dagger}b_{k+qc} , \nn &&
j_{ba}^{x}(q,i\Omega) = - 4 t E X \sum_{k} \sin \Bigl(k_{x} +
\frac{q_{x}}{2}\Bigr) b_{kc}^{\dagger}b_{k+qc} , \nn &&
K_{aA}^{bx}(q,i\Omega) = - 4 t E X \sum_{k} \cos\Bigl(k_{x} +
\frac{q_{x}}{2}\Bigr) b_{kc}^{\dagger}b_{k+qc} , \nn &&
j_{fa}^{x}(q,i\Omega) = 2 t F X \sum_{k} \sin \Bigl(k_{x} +
\frac{q_{x}}{2}\Bigr) \psi_{k\sigma}^{\dagger}\psi_{k+q\sigma} ,
\nn && K_{aa}^{fx}(q,i\Omega) = - 2 t F X \sum_{k} \cos\Bigl(
k_{x} + \frac{q_{x}}{2} \Bigr)
\psi_{k\sigma}^{\dagger}\psi_{k+q\sigma} . \nn   \eqa In this
expression we take the following replacement \bqa \cos \Bigl(
\phi_{ic} - \phi_{jc} \Bigr) && \rightarrow \frac{1}{2} \Bigl(
b_{ic}^{\dagger}b_{jc} + b_{jc}^{\dagger}b_{ic} \Bigr) , \nn \sin
\Bigl( \phi_{ic} - \phi_{jc} \Bigr) && \rightarrow \frac{i}{2}
\Bigl( b_{ic}^{\dagger}b_{jc} - b_{jc}^{\dagger}b_{ic} \Bigr)
\nonumber \eqa for evaluation of correlation functions.

Inserting Eq. (25) into Eq. (24), we find each current-current
correlation function in terms of each Green's function, \bqa &&
\pi^{b xx}_{AA}(q,i\Omega) = 16 t^{2} E^{2} Y^{2} \sum_{k}
\sin\Bigl(k_{x} + \frac{q_{x}}{2}\Bigr) \sin\Bigl(k_{x} -
\frac{q_{x}}{2}\Bigr) \nn && \frac{1}{\beta}\sum_{i\nu}
G_{b}(k+q,i\Omega+i\nu) G_{b}(k,i\nu) \nn && + 4 t E Y
\frac{1}{\beta}\sum_{i\nu}\sum_{k} \cos k_{x} G_{b}(k,i\nu)
\delta(q) \delta(i\Omega) , \nn && \pi^{b xx}_{aa}(q,i\Omega) = 16
t^{2} E^{2} X^{2} \sum_{k} \sin\Bigl(k_{x} + \frac{q_{x}}{2}\Bigr)
\sin\Bigl(k_{x} - \frac{q_{x}}{2}\Bigr) \nn &&
\frac{1}{\beta}\sum_{i\nu} G_{b}(k+q,i\Omega+i\nu) G_{b}(k,i\nu)
\nn && + 4 t E X \frac{1}{\beta}\sum_{i\nu}\sum_{k} \sin k_{x}
G_{b}(k,i\nu) \delta(q) \delta(i\Omega) , \nn && \pi^{b
xx}_{aA}(q,i\Omega) = 16 t^{2} E^{2} X Y \sum_{k} \sin\Bigl(k_{x}
+ \frac{q_{x}}{2}\Bigr) \sin\Bigl(k_{x} - \frac{q_{x}}{2}\Bigr)
\nn && \frac{1}{\beta}\sum_{i\nu} G_{b}(k+q,i\Omega+i\nu)
G_{b}(k,i\nu) \nn && + 4 t E X \frac{1}{\beta}\sum_{i\nu}\sum_{k}
\cos k_{x} G_{b}(k,i\nu) \delta(q) \delta(i\Omega) \eqa for pairon
excitations and
\bqa && \pi^{f}_{aa}(q,i\Omega) = - 4 t^{2} F^{2} X^{2} \sum_{k}
\sin\Bigl(k_{x} + \frac{q_{x}}{2}\Bigr) \sin\Bigl(k_{x} -
\frac{q_{x}}{2}\Bigr) \nn && \frac{1}{\beta}\sum_{i\omega}
\mathbf{tr} \Bigl\{ \mathbf{G}_{f}(k+q,i\Omega+i\omega)
\mathbf{G}_{f}(k,i\omega) \Bigr\} \nn && - 2 t F X
\frac{1}{\beta}\sum_{i\omega} \sum_{k} \cos k_{x} \mathbf{tr}
\Bigl\{ \tau_{z} \mathbf{G}_{f}(k,i\omega) \Bigr\} \delta(q)
\delta(i\Omega) \eqa for spinon excitations.

The pairon propagator is \bqa && G_{b}(q,i\Omega) = - b^{2}
\delta_{q,0}\delta_{\Omega,0} \nn && +
\frac{u}{\mathcal{E}_{q}^{b}} \Bigl[ \frac{1}{ i \Omega - \mu -
\mathcal{E}_{q}^{b}} - \frac{1}{ i \Omega - \mu +
\mathcal{E}_{q}^{b}} \Bigr] ,  \eqa and the spinon
Nambu-propagator is \bqa && \mathbf{G}_{f}(k,i\omega)
\equiv - \Bigl\langle \left( \begin{array}{c} \psi_{k\uparrow}  \\
\psi_{-k\downarrow}^{\dagger}
\end{array} \right) \left(
\begin{array}{cc} \psi_{k\uparrow}^{\dagger} & \psi_{-k\downarrow}
\end{array} \right) \Bigr\rangle \nn && \equiv \left(
\begin{array}{cc} G_{f}(k,i\omega) & F(k,i\omega) \\
F^{*}(k,i\omega)  & - G_{f}^{*}(k,i\omega)
\end{array} \right) , \nonumber \eqa where the normal Green's function is
\bqa && G_{f}(k,i\omega) = \frac{1}{2} \Bigl[ \frac{1 -
\frac{\varphi + 2 t F X \gamma_{k}}{E_{k}^{f}}}{i\omega -
E_{k}^{f}} + \frac{1 + \frac{\varphi + 2 t F X
\gamma_{k}}{E_{k}^{f}}}{i\omega + E_{k}^{f}} \Bigr] \nn \eqa and
the anomalous propagator \bqa && F(k,i\omega) = - \frac{t F Y
\varphi_{k}}{E_{k}^{f}} \Bigl[ \frac{1}{i\omega - E_{k}^{f}} -
\frac{1}{i\omega + E_{k}^{f}} \Bigr] . \eqa Inserting these
Green's functions into Eqs. (26) and (27), and performing the
Matsubara frequency summation, we obtain final expressions for all
current-current correlation functions in the static limit, \bqa &&
\pi^{b xx}_{AA}(q \rightarrow 0,i\Omega = 0) \nn && \approx 16
t^{2} E^{2} Y^{2} u^{2} \sum_{k} \frac{\sin^{2}
k_{x}}{\mathcal{E}_{k}^{b2}} \Bigl\{ \Bigl( - \frac{\partial
n(\mathcal{E}_{k}^{b} + \mu)}{
\partial [\mathcal{E}_{k}^{b} + \mu]} \Bigr) \nn && + \frac{n( \mathcal{E}_{k}^{b} + \mu)
- n(-\mathcal{E}_{k}^{b} + \mu)}{ \mathcal{E}_{k}^{b} } \Bigr\} -
4 t E Y b^{2} \nn &&  - 4 t E Y u \sum_{k} \frac{\cos
k_{x}}{\mathcal{E}_{k}^{b}} [n(\mathcal{E}_{k}^{b} + \mu) - n(-
\mathcal{E}_{k}^{b} + \mu)] , \nn && \pi^{b xx}_{aa}(q \rightarrow
0,i\Omega = 0) \nn && \approx 16 t^{2} E^{2} X^{2} u^{2} \sum_{k}
\frac{\sin^{2} k_{x}}{\mathcal{E}_{k}^{b2}} \Bigl\{ \Bigl( -
\frac{\partial n(\mathcal{E}_{k}^{b} + \mu)}{
\partial [\mathcal{E}_{k}^{b} + \mu]} \Bigr) \nn && + \frac{n( \mathcal{E}_{k}^{b} + \mu)
- n(-\mathcal{E}_{k}^{b} + \mu)}{ \mathcal{E}_{k}^{b} } \Bigr\}
\nn && - 4 t E X u \sum_{k} \frac{\sin k_{x}
}{\mathcal{E}_{k}^{b}} [ n(\mathcal{E}_{k}^{b} + \mu) - n(-
\mathcal{E}_{k}^{b} + \mu) ]  , \nn && \pi^{b
xx}_{aA}(q\rightarrow 0, i\Omega = 0) \nn && \approx  16 t^{2}
E^{2} X Y u^{2} \sum_{k} \frac{\sin^{2}
k_{x}}{\mathcal{E}_{k}^{b2}} \Bigl\{ \Bigl( - \frac{\partial
n(\mathcal{E}_{k}^{b} + \mu)}{
\partial [\mathcal{E}_{k}^{b} + \mu]} \Bigr) \nn && + \frac{n( \mathcal{E}_{k}^{b} + \mu)
- n(-\mathcal{E}_{k}^{b} + \mu)}{ \mathcal{E}_{k}^{b} } \Bigr\} -
4 t E X b^{2} \nn &&  - 4 t E X u \sum_{k} \frac{\cos k_{x}
}{\mathcal{E}_{k}^{b}} [ n(\mathcal{E}_{k}^{b} + \mu) - n(-
\mathcal{E}_{k}^{b} + \mu) ]  \eqa for pairons and \bqa &&
\pi^{f}_{aa}(q \rightarrow 0, i\Omega = 0) \nn && \approx - 4
t^{2} F^{2} X^{2} \sum_{k} \sin^{2} k_{x} \Bigl( - \frac{\partial
f(E_{k}^{f}) }{ \partial E_{k}^{f}} \Bigr) \nn && - 2 t F X
\sum_{k} \cos k_{x} \Bigl(\frac{ \varphi + 2 t F X
\gamma_{k}}{E_{k}^{f}} \Bigr) \tanh\Bigl( \frac{\beta
E_{k}^{f}}{2} \Bigr)   \eqa for spinons, where the spinon
contribution is basically the same as that of the BCS theory
\cite{dBCS}. Inserting Eqs. (31) and (32) into Eq. (23), we find
the superfluid density as a function of hole concentration and
temperature in the U(1) pair-rotor mean-field theory.

\subsection{Superfluid density as a function of hole concentration
and temperature}

We find that the dominant contribution is given by \bqa &&
\rho_{s}(T) \approx - \pi^{b xx}_{AA}(q \rightarrow 0,i\Omega =
0;T) \nn && = 4 t E Y b^{2} + 4 t E Y u \sum_{k} \frac{\cos
k_{x}}{\mathcal{E}_{k}^{b}} [n(\mathcal{E}_{k}^{b} + \mu) - n(-
\mathcal{E}_{k}^{b} + \mu)] \nn && - 16 t^{2} E^{2} Y^{2} u^{2}
\sum_{k} \frac{\sin^{2} k_{x}}{\mathcal{E}_{k}^{b2}} \Bigl\{
\Bigl( - \frac{\partial n(\mathcal{E}_{k}^{b} + \mu)}{
\partial [\mathcal{E}_{k}^{b} + \mu]} \Bigr) \nn && + \frac{n( \mathcal{E}_{k}^{b} + \mu)
- n(-\mathcal{E}_{k}^{b} + \mu)}{ \mathcal{E}_{k}^{b} } \Bigr\} ,
\eqa where the first two terms are diamagnetic contributions and
the last term is paramagnetic. This expression is simplified as
\bqa && \rho_{s}(T) \equiv \rho_{s} + \Bigl( \frac{d
\rho_{s}(T)}{d T} \Bigr)_{T \rightarrow 0} T , \eqa where the zero
temperature superfluid density is $\rho_{s} \approx 4 t E Y b^{2}$
and the decreasing ratio is $\Bigl( \frac{d \rho_{s}(T)}{d T}
\Bigr)_{T \rightarrow 0} \approx \frac{1}{\Lambda^{2}} \ln \Bigl(
- \frac{2\Lambda\sqrt{u t E Y}}{\mu}\Bigr)$ with momentum cutoff
$\Lambda$.

\begin{figure}[t]
\vspace{5cm} \includegraphics{SFXU_MD.eps} \caption{ (Color online) The
decreasing ratio of superfluid density $\rho_{s}(T)$ is enhanced
as hole concentration is reduced, where $\delta = 0.01$ (solid),
$\delta = 0.03$ (dashed), $\delta = 0.05$ (dash-dot), $\delta =
0.07$ (dash-dot-dot), and $\delta = 0.09$ (dotted). } \label{fig2}
\end{figure}

Fig. 2 shows the superfluid density with various hole doping.
Interestingly,  the decreasing ratio of the superfluid density is
enhanced as hole concentration is reduced, giving rise to the
monotonically increasing $T_{c}(\delta)$ in Fig. 1. We interpret
this tendency as the realization of the density-phase uncertainty
principle because phase fluctuations of Cooper pairs are stronger
in small doping.

We would like to point out that reduction of the superfluid
density originates from phase fluctuations in the U(1) pair-rotor
theory instead of scattering with Dirac fermions \cite{SF_Dirac}.
The contributions of Dirac fermions also result in the
temperature-linear decreasing ratio. However, such contributions
become irrelevant in the Ioffe-Larkin expression, resulting from
non-minimal coupling to gauge fields in the pairon sector.

\section{Discussion}

\subsection{Comparison with the BCS theory}

It is interesting to observe that the U(1) pair-rotor theory [Eq.
(14)] is almost "dual" to the slave-fermion theory \cite{Jia_Kim},
where the charge SU(2) symmetry is replaced with the spin SU(2)
symmetry. In this respect the pseudogap state, where Cooper pairs
are not coherent globally, is a mirror image of the so called
anomalous metal phase, sometimes referred as the algebraic charge
liquid in the slave-fermion description
\cite{Jia_Kim,ACL,Kim_Kim}, where antiferromagnetic correlations
exist only locally. Emergence of such an anomalous state reflects
strong quantum fluctuations, based on the uncertainty principle.

It is important to notice that the SU(2) slave-rotor
representation is difficult to be applied to the
negative-interaction model because the time-fluctuation term in
the rotor Lagrangian of Eq. (2) becomes negative, thus its
path-integral representation is not defined consistently. This
implies that the U(1) pair-rotor theory differs from the BCS
theory in itself.

\subsection{Origin of spectral asymmetry}

We show that the spectral asymmetry \cite{Asymmetry} appears
naturally in the U(1) pair-rotor theory. The electron Green's
function is given by multiplication of the boson [Eq. (28)] and
fermion [Eqs. (29) and (30)] Green's functions, $G_{xx'}^{el}
\approx - G_{xx'}^{b} \Bigl( G_{xx'}^{f} + F_{xx'}^{*} + F_{xx'} -
G_{xx'}^{f*} \Bigr)$, where $F_{xx'}$ is an anomalous Green's
function due to pairing. Then, we obtain the spectral intensity
$A_{el}(k,\omega) = A_{coh}(k,\omega) + A_{in}(k,\omega)$, where
$A_{in}(k,\omega)$ is an incoherent background and the coherent
part is \bqa && A_{coh}(k,\omega) = b^{2} \Bigl\{ \Bigl( 1 -
\frac{2 t F Y \varphi_{k}}{E_{k}^{f}} \Bigr) \delta(\omega -
E_{k}^{f}) \nn && + \Bigl( 1 + \frac{2 t F Y
\varphi_{k}}{E_{k}^{f}} \Bigr) \delta(\omega + E_{k}^{f}) \Bigr\}
,  \eqa showing the spectral asymmetry which originates from
pairing correlations. This predicts that the spectral asymmetry
will disappear when pairing correlations vanish at a temperature,
usually identified with the pseudogap temperature $T^{*}$. More
quantitative analysis is necessary.

\subsection{Application of the SU(2) slave-rotor theory to one
dimension}

It is valuable to apply the SU(2) slave-rotor representation to
the one-dimensional Hubbard model. Actually, one has the same
SU(2) slave-rotor Lagrangian as Eq. (2) in one dimension.
Considering that the fermion part is an SU(2) gauge theory in
$(1+1)D$, non-Abelian bosonization of QCD$_{2}$ results in the
SU$_{k=1}$(2) WZNW (Wess-Zumino-Novikov-Witten) theory
\cite{Tsvelik} with level $k$. Combining this fermion sector with
the pairon part, we obtain the SO(4) WZNW theory for spin dynamics
and SO(4) nonlinear $\sigma$ model without the topological term
for charge dynamics at half filling ($\mu = 0$), where spin
dynamics is decoupled from charge dynamics \cite{Tsvelik}. As a
result, charge fluctuations are gapped, corresponding to a Mott
insulator, while spin excitations are critical due to the presence
of the topological term. Hence charge fluctuations are described
by the pair-rotor Lagrangian even in one dimension, implying that
our formulation generalizes the bosonization scheme of one
dimensional charge dynamics.

\section{Conclusion}

In this paper we proposed a mechanism of superconductivity based
on the U(1) pair-rotor theory, where quantum fluctuations for
phase dynamics of Cooper pairs are taken into account. An
important feature is that the single energy scale for the Cooper
pair formation and phase coherence is separated into two energy
scales, allowing the pseudogap phase, where quantum phase
fluctuations are so strong as to destroy the superconductivity,
but superconducting correlations still exist at least locally. We
argued that emergence of such two energy scales is the hallmark of
the strong coupling approach as the Schwinger-boson theory for
antiferromagnetism of localized spins where the spin-gap phase
corresponds to the pseudogap state, differentiated from the weak
coupling approach such as the BCS theory \cite{Eliashberg} or HMM
framework \cite{HMM}.

\acknowledgements

K.-S. Kim thanks J.-H. Han for helpful discussions. K.-S. Kim was
supported by the National Research Foundation of Korea (NRF) grant
funded by the Korea government (MEST) (No. 2009-0074542).

\end{document}